\documentclass[12pt]{article}
\usepackage{amsfonts,epsfig}
\usepackage{amsmath,amsfonts}
\usepackage{amssymb,graphicx}

\def\pa{\partial}

\def\s{\sigma}

\def\p{\varphi}
\def\k{\kappa}

\def\d{\delta}
\def\r{\rho}
\def\h{\hat}

\def\l{\label}
\def\a{\alpha}
\def\bt{\beta}
\def\la{\lambda}
\def\g{\gamma}
\def\m{\mu}
\def\n{\nu}
\def\be{\begin{equation}}
\def\ee{\end{equation}}
\def\ba{\begin{eqnarray}}
\def\ea{\end{eqnarray}}

\def\h{\hat}
\def\b{\bar}
\def\t{\tilde}
\def\f{\frac}
\def\G{\Gamma}
\def\e{\epsilon}

\begin{document}

\vspace{4cm}

\begin{center}{\Large\bf 
Bounce of gravitons emitted from the brane to the bulk back to the brane.}
\end{center}
\vspace{1cm}
\centerline{\bf Mikhail Z. Iofa \footnote{ e-mail:
iofa@theory.sinp.msu.ru}} \centerline{Skobeltsyn Institute of Nuclear
Physics}
\centerline{Moscow State University}
\centerline{Moscow 119992, Russia}
\vspace{1 cm}
\begin{abstract}
Solutions of geodesic equations describing propagation of gravitons in the bulk
are studied in a cosmological model with one extra dimension. Brane with matter is
embedded in the bulk.
It is shown that in the period of early cosmology gravitons emitted from the brane to the
bulk under certain conditions can
return back to the brane.
The model is discussed in two alternative approaches:
(i) brane with  static metric moving  in the AdS space,
and (ii)  brane located at a fixed position in extra dimension  with  non-static metric.
Transformation of coordinates from the one picture to
another is performed.
In both approaches
conditions for gravitons emitted to the bulk to come back to the brane are found.

\end{abstract}
{\it Keywords}: extra dimensions; graviton emission; geodesic equations  
\vspace{0.5cm}

\section{Introduction}
Models of the space-time with extra dimensions beyond the four 
macroscopic
can be considered as candidates for alternative cosmologies. The minimal 
requirement on these models is  an ability to reproduce the main 
observational cosmological data such as the age of the Universe, abundance of light 
elements produced in primordial nucleosynthesis, etc.

In this paper we study a 5D model with one 
infinite extra dimension with the 3-brane with matter embedded in the bulk
\cite {BDL1,BDL2,maart}. 
In the 
radiation-dominated period of the Universe interacting particles of hot 
matter on the brane produce graviton 
emission which can escape to the bulk. 
This effect alters time evolution of matter on the brane as 
compared to the Standard cosmological model, and, in particular, alters  primordial 
nucleosynthesis \cite{heb,lan2,lan1,mi}. 
We study solutions of geodesic equations for gravitons propagating in the bulk.
In the period of early cosmology, in which in the generalized Friedmann equation  the term quadratic 
in brane energy density is dominant, there appears a new effect: gravitons emitted in the 
bulk can bounce back to the brane.

In a class of metrics defined below, the 5D model can be treated in two alternative 
approaches. 
In the first approach the brane is located at a fixed position
in the extra dimension, and the metric is time-dependent. 
In the second approach the brane is moving in the static 5D
space-time, constructed by attaching  two AdS spaces
at both sides of the brane. We consider a method to connect both approaches (cf. \cite{bow,muk,wu}).

Solving the geodesic equations in the period of early cosmology, 
in both approaches,
we find solutions for trajectories of gravitons propagating in the bulk
and conditions for emitted gravitons 
 to bounce back to the brane. Related issues were discussed in \cite{ish,cald,abd}.

\section{Static and time-dependent netrics} 

In the first approach we consider a class of non-static metrics 
\cite{BDL1,BDL2}
\be
\l{a3}
ds^2 =- n^2  (y,t) dt^2 +  a^2 (y,t) dx^a dx_a +dy^2 .
\ee
The components of the metric $n^2 (y,t)$ and $a^2 (y,t)$ satisfy  the
system of 5D Einstein
equations with the energy-momentum tensor of matter confined to the brane  and
the cosmological constant $\Lambda$.
The brane is located at a fixed position in
the extra coordinate, which we set $y=0$.
The function
$n(y,t)$ is normalized by
condition $ n(0,t) =1$.
For simplicity we consider a spatially flat brane.

The energy-momentum tensor on the brane is
\be
\l{em}
\tau_\m^\n =diag\{-\h{\r}-\h{\s}, \h{p}-\h{\s},\h{p}-\h{\s},\h{p}-\h{\s}\},
\ee
where $\h{\r}$ and $\h{p}$ are the energy and momentum densities of
matter on the brane, $\h{\s}$ is the tension of the brane.

It is convenient to define the normalized energy density, momentum and tension of the
brane
$$
\r =\f{\k^2\h{\r}}{6}, \qquad\p =\f{\k^2\h{\p}}{6},\qquad \s =\f{\k^2\h{\s}}{6},
$$
where $\h{\r}$ and $\h{\s}$ have dimension $[mass]^4$ and $\r$ and $\s$
have dimension $[mass]^1$.
Parameter $\m$ having dimension $[mass]$ is defined via  the 5D
cosmological constant $\m^2 =\k^2 |\Lambda|/6$, where $\k^2 =8\pi/M^3$ is
the 5D gravitational coupling constant.

In the
class of metrics (\ref{a3}) there is a unique extension of the metric
from the brane to the bulk \cite{BDL2}
\ba
\l{a4}
&{}&a^2 (y,t) =\f{a^2 (0,t)}{4}\left[e^{2\m |y|}
\left(\left(\f{\r +\s}{\m}-1 \right)^2 +\f{\r_w}{\m}\right)+
\right.\\\nonumber &{}& \left.
e^{-2\m |y|} \left(\left(\f{\r +\s}{\m}+1 \right)^2
+\f{\r_w}{\m}\right)
-2\left(\left(\f{\r +\s}{\m}\right)^2 -1
+\f{\r_w}{\m}\right)\right]
,\ea
and $ n(y,t) =\dot{a}(y,t)/\dot{a}(0,t)$.
The function $ a(0,t)$ satisfies the generalized Friedmann
equation (\ref{a2}) \cite{BDL1,BDL2}
\be
\l{a21}
H^2 (t) = -\m^2 +(\r +\s )^2 +\m \r_w (t)
,\ee
where 
$$
H(t) =\f{\dot{a}(0,t)}{a(0,t)},\qquad  \r_w (t) =\f{ \r_{w0}}{a^4 (0,t)}.
$$
are the Hubble function and the Weyl radiation term.
In derivation of (\ref{a21}) $\r_{w0}$ appears as an integration constant \cite{BDL2}. 

In the second approach the moving brane separates 
two static 5D AdS spaces attached at both sides of the brane with the metrics 
\cite{kraus,col,cham,bir}
\be
\l{a1}
ds^2 = -f(R) dT^2 +\f{d R^2}{f(R)} +\m^2 R^2 dx^a dx_a \equiv g_{ij}dx^i dx^j
.\ee
Here
$$
f(R)= \m^2 R^2 - \f{P}{R^2}
.$$
$R$ and $T$ have dimension $[mass]^{-1}$. 
Reduction of  the metric on the brane is $ds^2 =-dt^2 +\m^2  R_b^2 (t) dx^a dx_a$, where $t$ is the proper 
time on the brane.

In a parametric form trajectory of the brane is defined through the proper time $t$ on the 
brane as
$R= R_b (t),\,\,T= T_b (t),\,\,x^a =0$, where $ -f(R_b )\dot{T}_b^2 
+f^{-1}(R_b )\dot{R}_b^2 =-1 $. It follows that
\be
\l{t}
\dot{T}_b=\xi \frac{\sqrt{f(R_b )+\dot{R}_b^2}}{f(R_b )},
\ee
where $\xi =\pm 1$.

From the spatial components ($a,b=1,2,3$) of the junction conditions on 
the brane 
follows the generalized Friedmann equation 
\cite{BDL1,BDL2,kraus,col,cham,maeda,maart} 
\be
\l{a2}
\left(\f{\dot{R_b}}{R_b}\right)^2 =-\m^2 +(\r +\s )^2 +\f{ P}{R_b^4} 
.\ee
Equation (\ref{a2}) is  of the same form as (\ref{a21}).
The terms $ a^2 (0,t)$ and  $\r_w (t) = \r_{w0}/a^4 (t)$ can be identified with the terms 
$ \m^2  R_b^2 (t)$ and
$P/\m R_b^4 (t)$. Below we set $\s =\m$ and consider the case with $P=0$
\cite{mi}. 

\section{Connection between two pictures}

To establish connection between two pictures, we consider geodesic equations.
In the picture with the static metric,  geodesic equations with 
parameter  $y$ along the geodesics  are
\ba
\l{a5}
&{}&\f{d^2 T}{dy^2}+2\G^T_{TR}\f{dT}{dy}\f{dR}{dy}=0
\\\l{1g3}
&{}&\f{d^2 x^a}{dy^2}+ 2\G^a_{bR}\f{dx^b}{dy}\f{dR}{dy}=0
\\\l{a7}
&{}&\f{d^2 R}{dy^2}+ \G^R_{RR}\left(\f{dR}{dy}\right)^2 +
\G^R_{TT}\left(\f{dT}{dy}\right)^2 
+\G^R_{ab}\f{dx^a}{dy}\f{dx^b}{dy}=0,
\ea
where the Christoffel symbols  are 
$$
\G^T_{TR}=\f{f'}{2f},\quad \G^R_{RR}=-\f{f'}{2f},\quad
\G^R_{TT}=\f{1}{2}ff' ,\quad \G^R_{ab} =-\eta_{ab}f\m^2 R,\quad
\G^a_{Rb}=\f{\d^a_b}{R}
.$$
Here $(T,\, R)\equiv (T^\pm ,\, R^\pm )$ are coordinates in the AdS spaces 
at the opposite sides of the brane $f=\m^2 R^2$.
Integrating the geodesic equations, one  obtains
\be
\l{a8}
\f{dT^\pm}{dy}=\f{E^\pm}{f(R)},\qquad \f{dx^a}{dy}=\f{C^a}{\m^2 R^2}, 
\qquad
\left(\f{dR^\pm}{dy}\right)^2 =f(R) ({C^{R\,\pm}})^2 +{E^{\pm}}^2 -{C^a}^2    
,\ee
where $ (E^\pm,\,C^{a} , \,C^{R\,\pm} ) $ are integration parameters.
$(dT/dy,\,dR/dy,\,dx^a /dy)$ are the components of the tangent vectors to 
geodesics which we normalize to unity.
Imposing the normalization condition
$$
\f{dx^i}{dy}\f{dx^j}{dy}g_{ij} =1, \qquad i=T,R,a
$$
we obtain that $({C^{R\,\pm}})^2 =1$.


Connection between the two pictures is established in the following way.
We consider solutions of the geodesic equations even in $y$:
$T^+ (y)=T^- (-y),\,\, R^+ (y)=R^- (-y)$ . 

The hypersurface $(T,\,R,\,x^a =0)$ is foliated by  geodesics that intersect 
trajectory of the
brane $(T_b (t),\,R_b (t),\,0)$ and at the intersection point are orthogonal to it (cf \cite{muk}).
Geodesics are labeled by parameter $t$.
$t$ is 
constant along a geodesic. 
The geodesics orthogonal to the trajectory of the brane satisfy
boundary  conditions at $y=0:\quad R^\pm (0,t)= R_b (t),\,T^\pm (0,t)= T_b (t)$.   
Setting $C^a =0$, have
\be
\l{a9}
\f{\pa T^{\pm}(y,t)}{\pa y}=\f{E\e (y)}{f(R)},
\qquad
\f{\pa R^{\pm}(y,t)}{\pa y} =\a\e (y)\left(f(R) +{E}^2 \right)^{1/2}
,\ee
where $\a =\pm 1$, and $E^+ =E^- =E$.

 The normalized velocity vector of the brane and 
the normal vector to the brane trajectory are
\be
\l{c6}
v^i_b =(\dot{T}_b ,\,\dot{R}_b) =\left(\xi\sqrt{f(R_b )+\dot{R}_b^2}/{f(R_b 
)},\,\dot{R}_b)\right)\qquad
n^{i\pm}_b =\eta\e(y) \left(\f{\dot{R}_b}{f(R_b )} ,\,\xi\sqrt{f(R_b )+\dot{R}_b^2} \right)
,\ee
where  $\eta =\pm 1$. 
From (\ref{a9}) we obtain the tangent vector to a geodesic
\footnote{On the brane we find $n^R =(\pa R/\pa y) n^y$ and $n^T =(\pa T/\pa y) n^y$, where $n^y$ is the normal 
to 
the brane in the metric (\ref{a3}). The same relations hold for $v^i_b$.} 
\be
\l{c1}
u^{i\pm} =\left(\f{E\e (y)}{f(R)},\,\,
\a\e (y)\sqrt{f(R)+E^2}\right)
.\ee
By construction, the tangent vector to a geodesic at the intersection point with
trajectory of the brane is (anti)parallel to the normal to the trajectory of the brane,
\be
\l{c1a}
u^i\,|_{y=0} || n^i_b
.\ee
From this condition it follows that
\be
\l{c7}
E=\eta \dot{R}_b ,\qquad \a=\xi\eta 
\ee

\section{Integration of geodesic equations}

Let us integrate the geodesic equations.
From the second equation (\ref{a9})  we obtain
\be
\l{10b}
R^\pm (y,t)=
R_b (t)\cosh\m y +\a\sqrt{\f{\dot{R}_b^2}{\m^2} +R_b^2}\,\sinh\m |y|
.\ee
 Using the Friedmann equation (\ref{a2}) with $P=0$
\be
\l{fr} 
H^2 =\dot{R}_b^2 /R^2_b =\r^2 +2\m\r
\ee
 and introducing $\bt =sign (R_b (t))$, we rewrite (\ref{10b}) as
\be
\l{110}
R^\pm (y,t) =R_b (t)\left(\cosh\m y +\a\bt \left (1 +\f {\r}{\m}\right)\sinh \m |y|\right)
.\ee
For correspondence with (\ref{a4}) (the case with $P=0$)
$$
a^2 (y,t)=\f{a^2 (0,t)}{4}\left[e^{2\m|y|}\left(\f{\r}{\m}\right)^2
+e^{-2\m|y|}\left(\f{\r}{\m}+2\right)^2 -2\f{\r}{\m}\left(\f{\r}{\m}+2\right)\right]
$$
we set 
$$
\a\bt =-1.
$$
 Omitting in (\ref{10b}) $(\pm)$, we obtain
\be
\l{111}
R(y,t) =\f{R_b (t)}{2}\left[e^{-\m |y|}\left(\f{\r}{\m}+2\right)-e^{\m |y|}\f{\r}{\m}\right]. 
\ee
Introducing $y_0$, such that
\be
\l{113}
e^{\m y_0}=\left(\frac{\r}{\r +2\m}\right)^{1/2}
\ee
we express $R(y,t)$ and $R' (y,t)$ as
\be
\l{115}
R(y,t)=-\f{\dot{R}_b (t)}{\m}\sinh (\m |y| +\m y_0 ),\qquad R' (y,t)=-\e (y) \dot{R}_b (t)\cosh (\m |y| 
+\m y_0 )
\ee
Integrating Eq. (\ref{a9}) for $T(y,t)$ with $R(y,t)$ (\ref{111}), we obtain
\be
\l{141}
T^\pm (y,t) =-\f{1}{\m E}\f{\cosh (\m |y| +\m y_0 )}{\sinh (\m |y| +\m y_0 )}
+C^\pm (t)
.\ee
For $C^+ (t) =C^- (t)=C(t)$ the limits $y=0$ from both sides of the trajectory of the brane coincide.

To determine $C(t)$, first, we consider transformation of the metric (\ref{a1}) with $P=0$ from  coordinates 
$R,\, T$ to $y,\,t$. We have
\be
\l{m1}
ds^2 =dy^2 \left(-\m^2 R^2 {T'}^2 +\f{{R'}^2}{\m^2 R^2}\right) 
+2dtdy\left(-\m^2 R^2\dot{T}T'+\f{\dot{R}R'}{\m^2 R^2}\right)
+dt^2\left(-\m^2 R^2 \dot{T}^2 +\f{\dot{R}^2}{\m^2 R^2}\right) +\m^2 R^2 {dx^a}^2
.\ee
Substituting $T'$ and $R'$  (\ref{a9}), we find that the coefficient at $dy^2$ is ${\e^2 (y)}$. 
The coefficient at $dy dt$ is zero, if
\be
\l{m2}
\dot{T} =\f{\dot{R}\,R'}{\m^4 R^4 \,T'}= \f{\dot{R}\,R'}{\m^2 R^2 (\e (y)E)},
\ee
where  in the second equality we have used (\ref{a9}).
Substituting  in the coefficient at $dt^2$ expression for $\dot{T}$ (\ref{m2}) and using 
(\ref{a9}) and (\ref{115}), we obtain 
${\dot{R}}^2/{\dot{R_b}}^2 =n^2 (y,t)$ (cf. (\ref{t})).

Writing (\ref{141}) as
\be
\l{m3}
T(y,t)=-\f{ R' (y,t)/\e (y)}{\m^2 R(y,t) E} +C(t)
\ee
and taking the time derivative, we obtain
\be
\l{m4}
\dot{T}= \f{\dot{R} R'}{\m^2 R^2 (\e (y) E )}-\f{1}{\m^2 R}\f{d}{dt}{\left(\f{R' \,\e (y)}{ 
E}\right)} 
+\dot{C}
\ee
The first term in the rhs of (\ref{m4}) is the same as in (\ref{m2}).
Remarkably, substituting explicit expressions (\ref{115}) for $R$ and $R'$, we find that the second term in 
the rhs of (\ref{m4}) is independent 
of $y$ 
\be
\l{m6}
\f{1}{\m^2 R}\f{d}{dt}{\left(\f{R' \,\e (y)}{ E}\right)}=\f{\eta \dot{y}_0 (t)}{\dot{R}_b (t)}
.\ee
Choosing
\be
\l{m8}
\dot{C}=\eta \f{\dot{y}_0}{ \dot{R}_b}
,\ee
 we obtain  $\dot{T}$ as in (\ref{m2}). In the limit $y=0$ we have
\footnote{Conversely, requiring that the limit $y=0$ of (\ref{m2}) is $\dot{T}_b (t)$, we arrive 
at (\ref{m8}).}
\be
\l{m7}
\dot{T}(y,t)|_{y\rightarrow 0}=-\eta\f{\r +\m }{\m^2 R_b}=\dot{T}_b (t)
.\ee
In the radiation-dominated period, conservation equation for the energy density on the brane is
$$
\dot{\r}=-4\r \dot{R}_b /R_b.
$$
Using this equation, we have 
$$
 \dot{y}_0 
=-\f{4\r}{H},\qquad \dot{T}_b =-\eta\f{\r +\m }{\m^2 R_b}
.$$
From (\ref{111}), using the conservation equation for $\r$, we obtain
\be
\l{145}
\dot{R}(y,t)=
\f{\dot{R}_b}{2}\left[e^{\m |y|}\f{3\r}{\m}-e^{-\m
|y|}\left(\f{3\r}{\m}-2\right)\right]
.\ee
Jacobian of transformation from $T,\,R$ to $t,\, y$ is 
$$
J= \dot{T}R'-\dot{R}T' =\e\eta\f{\dot{R}}{\dot{R}_b}.
$$
Although both $\dot{R}$ and $\dot{R}_b$ are non-zero, the expressions for 
$\dot{T}$ and $T'$ contain $R(y,t)$ in denominator. $R(y,t)$ is zero for $e^{2\m |y|} =(\r +2\m 
)/\r $, 
 and transformation from $(T,\,R)$ to $(t,\, y)$ 
is valid for $e^{2\m |y|}< 1+2\m/\r$. 
In the region $e^{2\m |y|}> 1+2\m/\r \,\,R(y,t)$ can be defined as 
$$
R(y,t) =\f{R_b (t)}{2}\left[e^{\m |y|}\f{\r}{\m}-e^{-\m |y|}\left(\f{\r}{\m}+2\right) \right]
,$$
which is positive and increasing function $y$.


\section{Null geodesics}

Next, we  consider null geodesics along which gravitons propagate in the bulk.
In the picture with the static space-time null geodesics again are found from 
Eqs. (\ref{a5})-(\ref{a7}). 
The tangent vectors to a null geodesic satisfy the relation
\be
\l{1g}
g_{ij}\f{dx^i}{d\la}\f{dx^j}{d\la}=0
,\ee
where, to distinguish the case of null geodesics, we relabeled the affine parameter 
from $y$ to $\la$.
Substituting solution (\ref{a8})  in (\ref{1g}), we obtain that  $C^R =0$. 
Taking for definiteness direction $\la>0$, we obtain tangent vectors to a null geodesic  
\be
\l{lg5}
\f{d{T}}{d\la}=\f{C^T}{f({R})},\qquad 
\f{d{x}^a}{d\la}=\f{C^a}{\m^2 {R}^2},
\qquad
\f{d{R}}{d\la} =
\nu |C^T | \sqrt{ (1-\g )}
,\ee
where  $\n =\pm 1$. 
Here 
$$
\g ={C^a}^2/{C^T}^2
.$$
 The geodesic equations can be integrated in a closed form.
\be
\l{lg6}
{R}(\la ,t )-{R}(0, t ) =\sqrt{{C^T}^2 (1-\g )}  \la  ,\qquad {T}(\la ,t
)-{T}(0, t )
=\f{C^T}{\m^2 {R}(0, t )}
\,\f{\la}{\sqrt{{C^T}^2 (1-\g )}   + {R}(0, t )}
\ee
Here $t$ is  proper time on the brane at which begins the geodesic.

Let $(R(\la ), \,T(\la ), \,x^a (\la ))$ is a point on a null geodesic. $R,\, T$ can be expressed as
functions of $y,\, t$ on the hyperplane $x^a =0$. For a point on a geodesic we have
$R(\la )=R(y(\la ), t(\la )),\,\,T(\la )=T(y(\la ), t(\la ))$.

\vspace{6cm}
\begin{figure}[ht]
\vspace*{-0.2cm}
\begin{picture}(100,100)
\thicklines
\put(100,25){\epsfig{file =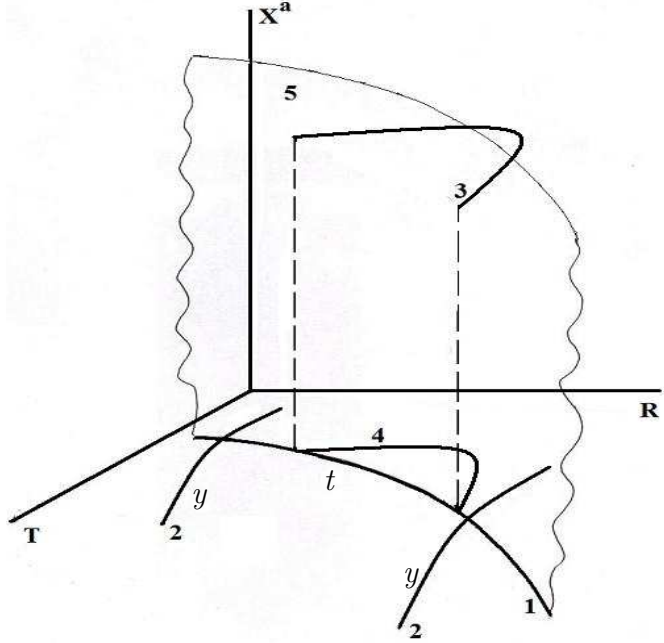 ,width =250pt, height
=250pt}}
\put(170,80){\it{y}}
\put(220,85){\it{t}}
\put(250,50){\it{y}}
\end{picture}
\vspace{-1.3cm}
\caption{Schematic plot of geodesics on the brane and in the bulk. Curve 1 is the intersection of the brane 
world scheet 5 with the hypersurface $R,\,T$ (brane trajectory).  Curves 2 represent the geodesics  
orthogonal to brane trajectory 
in the  hypersurface $R,\,T$. At the brane trajectory $y=0$, and $t$ is running along trajectory. At the 
geodesics orthogonal to the brane trajectory $t=const$ and $y$ varies along geodesics.  
Curve 3 is a null geodesic in the bulk bouncing back to the brane. Curve 4 is projection of the null 
geodesic on the  hypersurface $R,\,T$.}
\vspace*{0.5cm}
\end{figure}
\vspace{-1cm}

Let us verify that the vector $(d{y(\la )}/d\la,\,d{t(\la )}/d\la,\,dx^a/d\la)$ is null in the 
time-dependent metric (\ref{a3})
\be
\l{nu}
\left(\f{d{y}}{d\la}\right)^2 -n^2 ({y},{t})\left(\f{d{t}}{d\la}\right)^2
+a^2 ({y},{t})\left(\f{d{x}^a}{d\la}\right)^2
\stackrel{?}{=}0,
\ee
where $a^2 (y,t) = \m^2 R^2 (y,t)$ and $n^2 (y,t) =\dot{R}^2 (y,t)/\dot{R}_b^2 (t)$.

From the equations
\ba
\l{rel}
&{}&\f{d {T}(\la )}{d\la}=\f{\pa {T}}{\pa t}\bigg|_{y}\f{d{t}}{d\la}+
\f{\pa {T}}{\pa y}\bigg|_{t} \f{d{y}}{d\la},
\\\nonumber
&{}&\f{d {R}(\la )}{d\la}=\f{\pa {R}}{\pa t}\bigg|_{y}\f{d{t}}{d\la}+
\f{\pa {R}}{\pa y}\bigg|_{t,} \f{d{y}}{d\la}
\ea
we obtain
\ba
\l{n1a}
\f{d{y}}{d\la}=\f{\dot{{R}}\,d{T}/d\la -\dot{{T}}\,d{R}/d\la}{\dot{{R}}\,{T}'
-\dot{{T}}\,{R}'},\\\nonumber
\f{d{t}}{d\la}=\f{{{T}'}\,d{R}/d\la -{{R}'}\,d{T}/d\la}{\dot{{R}}\,{T}'
-\dot{{T}}\,{R}'}
\ea
Substituting expressions (\ref{n1a}) in (\ref{nu}), we rewrite this relation as
\ba
\l{n1b}
&{}&\left(\f{d{R}}{d\la}\right)^2 \left(\dot{{T}}^2-\f{\dot{{R}}^2}{\dot{{R}}^2_b}{{{T'}}}^2\right)+
\left(\f{d{T}}{d\la}\right)^2
\left(\dot{{R}}^2-\f{\dot{{R}}^2}{\dot{{R}}^2_b}{{{R'}}}^2\right)\\\nonumber
&{}&
-2\f{d{R}}{d\la}\f{d{T}}{d\la}\left(\dot{{R}}\dot{{T}}
-\f{\dot{{R}}^2}{\dot{{R}}^2_b}{{{T}'}}{{{R}'}}\right)^2
+ \m^2 {R}^2({\dot{{R}}\,{T}'
-\dot{{T}}\,{R}'})^2\left(\f{d{x}^a}{d\la}\right)^2 \stackrel{?}{=}0
\ea
Using the  expressions (\ref{m2}) for $\dot{T}$, (\ref{111}) - (\ref{112}) for $R,R'$, (\ref{a9}) for  $T'$
and (\ref{lg5}) for  $(d\t{R}/d\la
,\,d\t{T}/d\la ,\, dx^a /d\la)$, we verify that the relation (\ref{n1b}) is valid.

In the 
setting with the non-static metric (\ref{a3}), 
the system of the geodesic equations is complicated, and it is difficult to 
integrate it in a closed form.
However, the equation for $x^a$ 
$$
\f{d^2 x^a}{d\la^2}+2\G^a_{ay}\f{dx^b}{d\la}\f{dy}{d\la}
+2\G^a_{bt}\f{dx^b}{d\la}\f{dt}{d\la}=0
$$
can be easily integrated and yields
$$
\f{dx^a}{d\la}=\f{C^a}{a^2 (y,t)}, 
$$
which is the same as in (\ref{lg5}).

\section{Conditions for bounce of gravitons}

Let us solve the combined system of equations for brane trajectory and a null geodesic and 
show that under certain conditions null geodesic can bounce back to the brane.
We consider the radiation-dominated period of the
early cosmology, when $\r >\m$. We assume that solving  Friedmann equation we can neglect the energy flow 
from the brane to the bulk  
\footnote{As it is seen from (\ref{b3}), it is sufficient to assume that the terms containing bulk
energy-momentum tensor  
are small. Expanding expression (\ref{b3}) in powers $\m/\r$, we obtain that in the leading order these 
terms are canceled.}.
We obtain
\be
\l{rho}
\f{\m}{\r (t)}= {4\m t +8(\m t)^2 },
\ee
and $H/\r = 1+\m t$.

As in \cite{lan2}, the equation for the brane trajectory can be written as
\be
\l{b3}
\f{dR_b}{dT_b}=\f{\m^2 R_b^2 \dot{R}_b}{\xi\sqrt{\m^2 R_b^2 +\dot{R}_b^2}}=\xi\bt\m^2
R_b^2 \f{H}{\sqrt{\m^2 +H^2}},
\ee
where  $\bt \equiv sign (R_b (t) )$.
Substituting $H^2 =\r^2 +2\m\r$ and expanding the rhs of (\ref{b3}) in powers of
$\m/\r <1$, we obtain
\be
\l{b31}
\f{dR_b}{dT_b}\simeq \xi\m^2 R_b^2 \left(1-\f{\m^2}{2\r^2 }\right).
.\ee
Using that $\r (t) =\r (t_1 )(R_b (t_1 )/R_b (t))^4$,  we have
$$
\left(1-\f{\m^2}{2\r^2 }\right)^{-1}\simeq 1+\f{\m^2 R_b^8 (t)}{2\r^2 (t_1 )R_b^8 (t_1 )}
.$$

Integrating Eq. (\ref{b31})  with the initial conditions $R_b =R_b (t_1 ),\,\, T_b =T_b
(t_1 )$, we have
\be
\l{b4}
\xi\bt\m^2 (T_b (t) -T_b (t_1 ))= -\f{1}{R_b (t)} +\f{1}{R_b (t_1)} +\f{\m^2}{14 (\r (t_1 
)R^4_b 
(t_1))^2}\left(R_b^7 (t) - R_b^7 (t_1 )\right)
.\ee
Here $\m^2/\r^2 (t_1 )$ is a small parameter.

Let us consider a graviton trajectory in the picture with the static metric.
From the equations (\ref{lg5}) we obtain  
\be
\l{b1}
\f{d\t{R}}{d\t{T}} =\n\k (1-\g )^{1/2}\m^2 \t{R}^2
,\ee
where $\k =sign (C^T )$.
Integrating this equation with the initial conditions $\t{R}=R_b (t_1 ),\,\, \t{T}=T_b 
(t_1 )$, 
we have
\be
\l{b2}
 \f{1}{R_b (t_1 )}-\f{1}{\t{R}} = \n\k (1-\g )^{1/2}\m^2 (\t{T}-T_b (t_1 ))
.\ee
To find, if at a time $t$ the graviton trajectory returns to the brane world sheet,
i.e. $\t{R}=R_b (t)$ and $\t{T}= T_b (t)$, we equate solutions (\ref{b4}) and (\ref{b2}).
In the case $\xi\bt =\n\k$ we have
\be
\l{b6}
[(1-\g )^{-1/2}-1]\left(\f{1}{R_b (t_1 )}-\f{1}{{R_b (t)}} \right) =
\f{\m^2}{14 \r^2 (t_1 )}\f{1}{R_b (t_1)}
\left(\left(\f{R_b (t)}{R_b (t_1 )}\right)^7 -1\right)
.\ee
Eq. (\ref{b6}) means that graviton emitted from the brane at time $t_1$ has bounced 
back on the brane at time $t$.

Using relations $\a\bt =-1$ and  $\a =\xi\eta$, condition $\xi\bt =\n\k$ is expressed as
\be
\l{b5}
\n\eta\k =-1
.\ee

Setting $z\equiv R_b (t_1 )/ R_b (t ) < 1$ and $\h{\g}/2\equiv [(1-\g )^{-1/2}-1]$, 
  we transform (\ref{b6}) as
\be
\l{11.9}
\f{(1-z^7)}{z^7 (1-z)}=\f{7\h{\g}\r^2 (t_1 )}{\m^2}
.\ee
The function of $z$ in the l.h.s. of (\ref{11.9}) is monotone decreasing with the minimum at $z=1$ equal 
to 7.
Eq. (\ref{11.9}) has unique 
solution, provided $\h{\g}\r^2 (t_1 )/\m^2 >1$.
Expressing $\r (t_1 )$ through  $\r (t )$, we obtain
\be
\l{11.10}
\f{z(1-z^7 )}{1-z} =\f{7\h{\g}\r^2 (t)}{\m^2}
.\ee
The function of $z$ in the l.h.s. of (\ref{11.10}) is monotone increasing with the maximum at $z=1$ equal
to 7.
Eq. (\ref{11.10}) has unique solution provided 
\be
\l{11.11}
\h{\g}\r^2 (t )/\m^2 <1.
\ee
Existence of solution means that graviton emitted from the brane at proper time $t_1$ comes 
back to the brane at time $t$.

Momentum of graviton is proportional to tangent vector to a geodesic (\ref{lg5}). In the 
basis $(e^A_T ,\,e^A_R ,\,e^A_a )$, where
$$
e^A_T = (1/\m R ,0,0,0,0),\quad e^A_R =(0,\m R,0,0,0),\quad e^A_a=(0,0,1_a/\m R ) 
$$
the lengths of momentum components are
$$
|p^T|^2=|g_{TT} {p^T}^2| ={C^T}^2/\m^2 R^2 ,\quad |p^R|^2={C^R}^2/\m^2 
R^2 ,\quad|p^a|^2={C^a}^2/\m^2 R^2 .
$$
In this basis condition (\ref{11.11}) means that $|p^T |\sim |p^R |\gg|p^a |$
\footnote{In different bases condition (\ref{11.11}) written in terms of components of 
momentum may take different form.}. 

\subsection{Bounce of gravitons in time-dependent metric}

Let us show how the bounce of gravitons takes place in the picture with the metric (\ref{a3}). The 
difficulty is that in this metric we have no analytic solution of geodesic equations.  We obtain  equation 
for graviton 
trajectory  using solution of geodesic equations (\ref{lg5}) and transforming it to coordinates $(t,\, y)$.

From the equations (\ref{n1a}) we obtain  the equation of trajectory 
\be
\l{n1}
\f{dy}{dt}=\f{\dot{R}\,dT/d\la -\dot{T}\,dR/d\la}{T'\,dR/d\la -R'\,dT/d\la}.
\ee
Using (\ref{lg5}), denominator in (\ref{n1}) is obtained as
\be
\l{n2}\nonumber
T'\,dR/d\la -R'\,dT/d\la =
\f{ C^T R_b \e (y)}{\m^2 R^2}\left[\nu\eta\k H(1-\g )^{1/2} +\f{(\r +2\m )e^{-\m |y|} +\r 
e^{\m |y|} }{2}\right]
.\ee
It is seen that the expression in brackets is positive.

The numerator can be written as
\be
\l{n3a}
\dot{R}\,dT/d\la -\dot{T}\,dR/d\la=
\f{C^T \dot{R}}{\m^2 R^2}\left(1+\nu\eta\k (1-\g )^{1/2}\f{e^{-\m |y|}(\r +2\m )+e^{\m 
|y|}\r}{2H}\right).
\ee
Eq. (\ref{n1}) takes the form
\ba
\l{n4}
\f{dz}{\m dt}=\nu\eta\k\f{1}{2}\left[z^2 \f{3\r}{\m} -\left(\f{3\r}{\m} -2\right)\right]
\f{2H +\n\eta\sqrt{1-\g}[z^{-1}(\r +2\m)+ z\r ]}
{2H \sqrt{1-\g} +\nu\eta\k [z^{-1}(\r +2\m)+ z\r ]}.
\ea
where $z=e^{\m|y|}$. 
In correspondence with (\ref{b5}) we take $\n\eta\k = -1$. 

The numerator has zeroes at the points
\be
\l{n5}
z_{1,2}(t)=\left(1+\f{2\m}{\r}\right)^{1/2}\left(\f{1\pm\sqrt{\g}}{1\mp\sqrt{\g}}\right)^{1/2}.
\ee
and is positive for $z_2<z<z_1$. 
At the initial time $t_1$   $y(t_1 )=0$ and $z(t_1 )=1$.            
Condition that $z_2 (t)\lessgtr 1$ is 
$$
\f{\m}{\m +\r (t )}\lessgtr \sqrt{\g}.
$$ 
The above formulas are valid  in the region where $R(y,t)>0$
$$
1<\b{z}(t)\equiv \left(1+\f{2\m}{\r}\right)^{1/2}.
$$ 
Eq. (\ref{n4}) can be written as
\be
\l{n6}
\f{dz}{\m dt}=-\f{1}{2}\left[z^2 \f{3\r}{\m} -\left(\f{3\r}{\m} -2\right)\right]
\f{(z-z_1 )(z-z_2 )}
{(z-z_1 )(z-z_2 )+2H\g z/\r\sqrt{1-\g}}
\sqrt{1-\g}.
\ee
Introducing new variable  $u=z-z_2 $, we have
\be
\l{n7}
\f{du}{\m dt}=-\f{1}{2}\left[ \f{3\r}{\m}((z_2 +u)^2 -1) +2)\right]
\f{-(z_1 -z_2 -u )u }{-u(z_1 -z_2 -u )+2\g \b{z}(z_2 +u)/\sqrt{1-\g }}\sqrt{1-\g }+\f{dz_2}{\m dt}.
\ee
Expanding all the terms in (\ref{n7}) to the linear order in $u$ and in small terms $\sqrt{\g}$ and $\m/\r$, 
we obtain
\be
\l{n9}
\f{du}{\m dt}=\f{1}{2}\left[\f{3\r}{\m}\left(\f{2\m}{\r}
-2\sqrt{\g}+2u\right)+2\right]
\f{u(u-2\b{z}\sqrt{\g})}
{u(u-2\b{z}\sqrt{\g})+2\g} 
-4\left(1-\f{\m}{\r}-\sqrt{\g}\right).
.\ee
At the initial time $t_1$ at which trajectory begins,  $\sqrt{\g}>\m/\r$ and  $z_2 (t_1 )<1$.  
 At larger times the point $z_2 (t)$ 
eventually
moves to the right and crosses unity at time $t_2$. At times $t>t_2$ we have $\sqrt{\g}<\m/\r$ 
and $z_2 (t)>1$.  
The initial time  $t_1$ at which trajectory begins can be chosen so that $z_2 (t_1 )$ is 
arbitrary close to 1: $1-z_2 (t_1 )=u(t_1 )\ll \sqrt{\g}$.
In this case $t_2 -t_1 $ is also small and $u(t_2)\ll \sqrt{\g}$.  

Eq. (\ref{n9}) is simplified as
\be
\l{n01}
\f{du}{\m dt}=\f{1}{2}\left[\f{3\r}{\m}\left(\f{2\m}{\r}-2\sqrt{\g}+2u\right)+2\right]
\f{u}{\sqrt{\g}-u} -4
=\f{3\r}{\m}\f{u^2 +u(8\m/3\r -\sqrt{\g})-4\m\sqrt{\g}/3\r}{\sqrt{\g}-u}.
\ee
As a function of $u$ the rhs of (\ref{n01}) is negative in the interval $(u_1,\,\,u_2 )$ and $\sqrt{\g}>u$,
$$
u_{1,2}=\f{1}{2}\left[\sqrt{\g}-\f{8\m}{3\r} \pm\sqrt{\left(\f{8\m}{3\r}\right)^2 +\g } \right].
$$
Here $u_1 >0$ and $u_2 <0$. 

 Eq. (\ref{n01}) can be written  as
$$
\f{du}{\m dt}=-\f{3\r}{\m}u -4\f{2u-\sqrt{\g}}{\sqrt{\g}-u}.
$$
For $u\ll \sqrt{\g}$, substituting $\r/\m\simeq 4\m t$, this equation can be simplified as 
\be
\l{n10}
\f{du}{\m dt}+\f{3u}{4\m t}=-4.
\ee
In the case  $\sqrt{\g}\gg u$  solution is
\be
\l{n11}
u(t)=\left(\f{t}{\b{t}}\right)^{-3/4}\left[u(\b{t})-\f{16}{7}\left(\f{t}{\b{t}}\right)^{7/4}\right]
,\ee
where $\b{t}$ is a reference time after which approximation is valid.
It is seen that $u({t})=0$ is reached in finite time $\b{t}$.
For larger times $t>\b{t}$ from 
(\ref{n6}) it follows that $ dz/\m 
dt<0$, i.e. the distance $z(t)$ from the brane decreases.

To conclude, in two approaches we have shown that under certain conditions emitted gravitons 
moving in the bulk along null geodesics can bounce back to the brane. 


I thank my colleagues at the Skobeltsyn Institute for discussion.
My special thanks are to I. Tuitin and B. Voronov and M. Smolyakov for useful remarks.


\end{document}